\documentclass{elsart}
\usepackage{graphicx,amssymb}
\journal{J. Phys. Chem. Solids}

\begin{document}

\begin{frontmatter}

\title{Magnetic behavior of a spin-1 dimer: \\ 
model system for homodinuclear \\ nickel(II) complexes}
\author[label1]{J. Stre\v{c}ka\corauthref{cor1}}, 
\ead{jozkos@pobox.sk}
\author[label1]{M. Ja\v{s}\v{c}ur},
\author[label2]{M. Hagiwara}, 
\author[label2]{Y. Narumi}, 
\author[label3]{J. Kuch\'ar},
\author[label2]{S. Kimura}, and
\author[label4]{K. Kindo}
\address[label1]{Department of Theoretical Physics and Astrophysics, 
Faculty of Science, \\ P. J. \v{S}af\'{a}rik University, Park Angelinum 9,
040 01 Ko\v{s}ice, Slovak Republic}
\address[label2]{KYOKUGEN (Research Center for Materials Science at Extreme Conditions), 
Osaka University, 1-3 Machikaneyama, Toyonaka, Osaka 560-8531, Japan}
\address[label3]{Department of Inorganic Chemistry, Faculty of Science, \\ 
P. J. \v{S}af\'{a}rik University, Moyzesova 11, 041 54 Ko\v{s}ice, Slovak Republic}
\address[label4]{Institute for Solid State Physics, University of Tokyo, \\ 
5-1-5 Kashiwa-no-ha, Kashiwa, Chiba 277-8581, Japan}
\corauth[cor1]{Corresponding author.} 
             
\begin{abstract}
Magnetic behavior of a spin-1 Heisenberg dimer is analysed in dependence on 
both uniaxial single-ion anisotropy and XXZ exchange anisotropy in a zero- as 
well as non-zero longitudinal magnetic field. A complete set of eigenfunctions and eigenvalues of the total Hamiltonian is presented together with an exact analytical expression for the Gibbs free energy, longitudinal magnetization, longitudinal and transverse susceptibility. The obtained theoretical results are compared with the relevant experimental data of [Ni$_2$(Medpt)$_2$($\mu$-ox)(H$_2$O)$_2$](ClO$_4$)$_2$.2H$_2$O 
(Medpt = methyl-bis(3-aminopropyl)amine).
\end{abstract}

\end{frontmatter}

\section{Introduction}

One of the most challenging tasks emerging in the field of molecular magnetism is 
to design molecule-based magnetic materials possessing predictable topology and desirable magnetic properties \cite{design}. A successfully realized rational synthesis of low-dimensional molecular magnets would provide,
among other matters, suitable model compounds for investigating magneto-structural correlations and in-depth understanding of cooperative quantum phenomena. Altogether, the simplest magnetic architecture among the molecular crystals have small {\it magnetic clusters} denoting the weakly-interacting assemblies of molecules formed by a few exchange-coupled paramagnetic centers \cite{cluster}.

It is of principal importance that a limited number of interacting sites in the discrete molecules often allows to model their physical properties within an exact quantum-mechanical approach that avoids a further approximation required for treating the extended solids. Among a large family of transition-metal coordination compounds are then {\it dinuclear complexes} the simplest model compounds which enable a precise description of quantum cooperativity and at the same time, provide an excellent testing ground for theories. Studies on 
their magnetic properties, moreover, played a crucial role in advancing the understanding of spin-exchange mechanism; it turned out that the nature and magnitude of exchange coupling depends on the ground-state electronic configuration of the constituting metal atoms as well as on the global topology and type of intervening non-magnetic atoms \cite{OK}. Besides, much effort directed towards the synthesis, design and analysis of dinuclear entities has been stimulated in connection with their wide application in various research fields. The metalloproteins with bimetallic cores often appear in the bioinorganic chemistry \cite{FO}, metalloenzymes (metalloproteins with a catalytic function) catalyse important chemical reactions at their metal centers in a variety of plants, fungi and bacterial species \cite{enzymes} and there is also a certain group of dinuclear complexes that shows an interesting photochemical behavior.

In the present work, we will focus our attention on modeling the magnetic behavior of spin-1 dimer with a particular emphasis aimed at better understanding of the influence of magnetic anisotropy. The main motivation to examine this model system can be related to a rich variety of existing homodinuclear nickel(II) complexes, which can serve as suitable model compounds for the system under investigation. The most successful strategy used for the preparation of dinuclear complexes consists in an appropriate combination of the {\it bridging} and ancillary {\it blocking} ligands. The former ones are capable of simultaneously binding two metal atoms in a close proximity and they also act as mediators of the superexchange interaction between the metal centers. By contrast, the latter ligands coordinatively saturate the metal centers 
in order to prevent the undesired self-assembly reaction providing the polymeric complexes. Under these circumstances, the dinickel cores in neighboring molecules are well-separated from each other and there is only weak inter-molecular interaction through the hydrogen bonds and/or van der Waals contacts. 

From the magnetic point of view, the dinuclear complexes can be classified according to the nature of 
exchange coupling and the type of bridging ligand. At present, there exists a large variety of dimeric compounds in which the nickel centers interact through the antiferromagnetic (AF), or ferromagnetic (FO) superexchange coupling. In spite of a remarkable diversity of bridging ligands emerging in the dinuclear nickel-based complexes, the series of {\it oxalato}- \cite{ox} and 1,3-{\it azido}-bridged \cite{az} compounds afford the generality of AF-coupled nickel dimers. On the other hand, almost all reported 
FO-coupled nickel dimers belong to the {\it halide}- or {\it pseudohalide}-bridged complexes \cite{halide} 
and among these, the coordination compounds with the 1,1-{\it azido} linkage provide their most extended sub-group \cite{fero}. It is worthy to notice that there also exists one rare exception of the FO dimeric compound, in which the dinickel core is embedded in the cavity of appropriate cryptand and the nickel atoms are linked via single azido group coordinated in the 1,3-bridging mode \cite{cryptate}. 

The organization of this paper is as follows. In Section 2, we will introduce the model system and we also recall the foundations for the occurrence of the magnetic anisotropy. This will be followed by a brief description of calculation procedure:
basic steps of derivation of the exact analytical relations for the Gibbs free energy, longitudinal magnetization, longitudinal and transverse susceptibility are listed along with a complete set of eigenfunctions and eigenvalues of the total Hamiltonian. The most interesting analytical and numerical results for the zero- and finite-temperature quantities are presented and discussed in Section 3.
A subsequent part, Section 4, serves for a comparison of obtained theoretical results with 
the relevant experimental measurements performed on one representative example of the spin-1 
dimeric compound. Finally, some concluding remarks are drawn in Section 5.
  
\section{Model system and its solution}

Before proceeding to an exact solution of the anisotropic spin-1 dimer, a few 
words should be addressed to a microscopic origin of the {\it magnetic anisotropy}. 
The global magnetic anisotropy of paramagnetic clusters is particularly associated 
with the single-ion anisotropy and/or the $g$-factor anisotropy which are combined 
with more or less isotropic exchange coupling between the interacting metal centers \cite{an}. It should be mentioned, however, that the main contribution to the magnetic anisotropy usually comes from the crystal field of ligands and hence, strongly anisotropic magnetic properties uprise just as the metal atoms reside in a crystal the sites of low (i.e. lower than cubic) symmetry. Consequently, the spatial anisotropy in response to an external magnetic field is vigorously reinforced by: a) {\it ligand asymmetry}, when different types of ligands coordinate to each metal atom; b) {\it geometric asymmetry}, when there is an inequivalent geometric arrangement of ligands around each metal atom; c) {\it coordination number asymmetry}, when the different number of ligands are coordinated to each metal atom \cite{ma}.

When the orbitally degenerate metal centers with a pronounced spin-orbit coupling build the metallic 
core of coordination compound, the exchange interaction between the magnetic ions becomes strongly anisotropic. This occurs for most of the rare-earth ions and also for some orbitally degenerate transition metal ions such as octahedral Co$^{2+}$ and low-spin Fe$^{3+}$ cations. It is worth noticing that the 
exchange anisotropy differs from the single-ion anisotropy as well as the anisotropy of $g$-factor. 
For example, the Yb$^{3+}$-- Cr$^{3+}$ exchange interaction in the YbCrBr$_9^{3-}$ 
face-sharing bioctahedral dimer is well-described by the extremely anisotropic spin Hamiltonian: 
$\hat {\mathcal H} = J_{xy} (\hat S_1^x \hat S_2^x + \hat S_1^y \hat S_2^y) 
                   + J_z \hat S_1^z \hat S_2^z$ with $J_{xy}$=-4.19cm$^{-1}$ 
and $J_{z}$=5.16cm$^{-1}$ despite both octahedrally coordinated Yb$^{3+}$ and Cr$^{3+}$ ions are by themselves magnetically isotropic \cite{miro1}. Several recent studies have shown that an obvious exchange anisotropy can be found in the compounds with other orbitally degenerate metal ions as well \cite{miro2}.

On the other hand, the individual magnetic centers with an orbitally non-dege\-ne\-rate ground state usually necessitate the isotropic exchange interaction between them, since the anisotropic term becomes due to orbital quenching just of minor importance. In such a case, the origin of exchange anisotropy lies merely in the admixture of excited states through the spin-orbit coupling and thus, it can be viewed only as a small higher-order perturbation to the isotropic exchange \cite{soc}. The ground state of Ni$^{2+}$ ion in an octahedral environment is, for instance, orbitally non-degenerate and as such, it implies more or less isotropic intra-dimer interaction in the \textit{homodinuclear nickel-based compounds}. Therefore, in an attempt to model the magnetic behavior of these compounds research has firstly focused on the theoretical interpretation based upon the assumption that the Ni$^{2+}$ ions are coupled by a purely isotropic interaction \cite{isotropic}. However, as it has been pointed out later on \cite{zfs}, the magnetic properties of octahedrally coordinated nickel compounds can be significantly affected by a rather large axial single-ion anisotropy. Bearing this in mind, Ginsberg {\it et al.} \cite{isotropicd} took into consideration the uniaxial zero-field splitting parameter regarding the influence of single-ion anisotropy as well. 

Although the exchange anisotropy is in generality negligible for the most 
of homodinuclear nickel(II) complexes, we will incorporate it here into the 
effective spin Hamiltonian to enable a deeper insight into the distinction 
between the effect of single-ion anisotropy and the effect of exchange anisotropy. 
To the best of our knowledge, such a comparative study has not been reported in the literature hitherto.
Now, let us write the total effective Hamiltonian of spin-1 dimer in the form: $\hat {\mathcal H} = \hat {\mathcal H}_{exch} + \hat {\mathcal H}_{azfs} + \hat {\mathcal H}_{zeet}$, which is composed of the exchange term, axial zero-field splitting term and Zeeman term, respectively, 
\begin{eqnarray}
&& \hat {\mathcal H}_{exch} = J_{xy} (\hat S_1^x \hat S_2^x + \hat S_1^y \hat S_2^y) 
                            + J_z \hat S_1^z \hat S_2^z, 
\label{eq1} \\
&& \hat {\mathcal H}_{azfs} = D [(\hat S_1^z)^2 + (\hat S_2^z)^2], 
\label{eq2} \\
&& \hat {\mathcal H}_{zeet} = - g_z \mu_{\mathrm{B}} B (\hat S_1^z + \hat S_2^z). 
\label{eq3}	   
\end{eqnarray}
Above, $\hat S_1^{\alpha}$ and $\hat S_2^{\alpha}$ ($\alpha$ = $x$, $y$, or $z$) 
denote the spatial components of the local spin-1 operator on the metal centers 1 
and 2, $J_{xy}$ and $J_z$ are the exchange couplings along the $x$, $y$ and $z$ directions, respectively, $D$ labels the axial zero-field splitting parameter and other quantities have an usual meaning: $B$ is the external magnetic field applied along the $z$-axis, $\mu_{\mathrm{B}}$ stands for Bohr magneton and $g_{\alpha}$ ($\alpha = x, y, z$) denotes the spatial component of $g$-factor. Notice that the negative (positive) sign of the zero-field splitting parameter $D$ corresponds to an easy-axis (easy-plane) single-ion
anisotropy. For easy reference, we will further refer to the $z$ axis as to the principal axis.

After performing elementary calculations in the usual matrix representation 
with a standard basis of functions $|S_1^z, S_2^z \rangle$ ($S_1^z = \pm 1,0$ and $S_2^z = \pm 1,0$), it is quite straightforward to find a complete set of eigenvalues and eigenfunctions of the total Hamiltonian: 
\begin{eqnarray}
&& \lambda_{0,0} = - \frac12 J_z + D - R, \quad  |\Psi_{0,0} \rangle = 
\frac12 [A_{+} (|1, -1 \rangle + |1, -1 \rangle) - \sqrt{2} A_{-} |0, 0 \rangle]; \nonumber \\
&& \lambda_{1, 0} = - J_{z} + 2D, \hspace{1.3cm}
|\Psi_{1, 0} \rangle = \frac{1}{\sqrt{2}} (|1, -1 \rangle - |-1, 1 \rangle); 
\nonumber \\
&& \lambda_{1, \pm 1} = - J_{xy} + D \mp H, \hspace{0.25cm}
|\Psi_{1, \pm 1} \rangle = \frac{1}{\sqrt{2}} (|\pm 1, 0 \rangle - |0, \pm 1 \rangle); 
\nonumber \\
&& \lambda_{2,0} = -\frac12 J_z + D + R, \quad |\Psi_{2,0} \rangle = \frac12 
    [A_{-} (|1, -1 \rangle + |1, -1 \rangle)+ \sqrt{2} A_{+} |0, 0 \rangle]; 
\nonumber \\
&& \lambda_{2, \pm 1} = J_{xy} + D \mp H,  \hspace{0.6cm}
|\Psi_{2, \pm 1} \rangle = \frac{1}{\sqrt{2}} (|\pm 1, 0 \rangle + |0, \pm 1 \rangle); \nonumber \\
&& \lambda_{2, \pm 2} = J_{z} + 2D \mp 2H, \hspace{0.4cm}
|\Psi_{2, \pm 2} \rangle = |\pm 1, \pm 1 \rangle.    
\label{eqe}	   
\end{eqnarray}
Here, we have introduced the expressions $H$, $R$ and $A_{\pm}$ in order to write 
the relevant eigenvalues and eigenvectors in a more abbreviated and elegant form: 
\begin{eqnarray}
H = g_z \mu_{\mathrm{B}} B; \quad
R = \sqrt{\Bigl( \frac{J_{z}}{2} - D \Bigr)^2 + 2 J_{xy}^2} \quad 
A_{\pm} = \sqrt{\frac{R \pm (\frac{J_z}{2} - D)}{R}}; . 
\label{eq4}	   
\end{eqnarray}

From here onward, one can readily proceed to the evaluation of basic 
thermodynamic quantities within the standard thermodynamical-statistical approach. 
Substituting the eigenvalues (\ref{eqe}) to a statistical definition of the partition function leads, for instance, to the following exact analytical expression:
\begin{eqnarray}
&& {\mathcal Z} = \sum_{\alpha} \exp(- \beta \lambda_{\alpha}) = 
   \exp[\beta (J_z - 2D)] + 2 \exp[- \beta (J_z + 2D)] \cosh(2 \beta H) \nonumber \\
             && + 4 \exp(- \beta D) \cosh(\beta H) \cosh(\beta J_{xy}) 
              + 2 \exp[\beta (J_z - 2D)/2] \cosh(\beta R),
\label{eq5}	   
\end{eqnarray}  
with $\beta = 1/k_{\mathrm{B}} T$. It should be stressed that the previous equation provides, because of the relation ${\mathcal G} = - \beta^{-1} \ln{\mathcal Z}$, an exact solution also for the Gibbs free energy ${\mathcal G}$. Hence, some other important quantities such as magnetization, pair correlation functions and quadrupolar moment can be in turn calculated by differentiating the Gibbs free energy  with respect to the relevant interaction parameter included in the total Hamiltonian. Thus, one easily obtains 
the magnetization per one site $m_z = - \frac12 \frac{\partial G}{\partial (\beta H)}$, pair correlation functions between the different spatial components of spins $C_{12}^{xx} = C_{12}^{yy} = \frac{\partial G}{\partial (\beta J_{xy})}$ and $C_{12}^{zz} = \frac{\partial G}{\partial (\beta J_{z})}$ at centers 1 and 2, or the $z$-component of quadrupolar moment $C_{11}^{zz} = \frac{\partial G}{\partial (\beta D)}$:
\begin{eqnarray}
m_z &=& \frac12 \langle \hat S_1^z + \hat S_2^z \rangle = \frac{1}{\mathcal Z} 
\Bigl \{ 2 \exp[- \beta (J_z + 2D)] \sinh(2 \beta H) \nonumber \\ && \qquad \qquad \qquad \qquad \qquad + 2 \exp(- \beta D) \sinh(\beta H) \cosh(\beta J_{xy}) \Bigr \}, 
\label{eq5a} \\
C_{12}^{xx} &=& \frac12 \langle \hat S_1^x \hat S_2^x + \hat S_1^y 
\hat S_2^y \rangle = - \frac{1}{\mathcal Z} \Bigl \{ 2 \exp(- \beta D) 
\cosh(\beta H) \sinh(\beta J_{xy}) \nonumber 
\\ && \qquad \qquad \qquad \qquad \qquad 
+ 2 \exp[\beta (J_z - 2D)/2] \sinh(\beta R)/R] \Bigr \}, 
\label{eq5b} \\
C_{12}^{zz} &=& \langle \hat S_1^z \hat S_2^z \rangle = \frac{1}{\mathcal Z} \Bigl \{ 
2 \exp[- \beta (J_z + 2D)] \cosh(2 \beta H) - \exp[\beta (J_z - 2D)] \nonumber \\
&& - \exp[\beta (J_z - 2D)/2] [\cosh(\beta R)+ (J_z - 2D) \sinh(\beta R)/2R] \Bigr \}, 
\label{eq5c} \\
C_{11}^{zz} &=& \frac12 \langle (\hat S_1^z)^2 + (\hat S_2^z)^2 \rangle = \frac{1}{\mathcal Z} \Bigl \{ 2 \exp[- \beta (J_z + 2D)] \cosh(2 \beta H) 
\nonumber \\ && \quad + 2 \exp(- \beta D) \cosh(\beta H) \cosh(\beta J_{xy}) 
+ \exp[\beta (J_z - 2D)] \nonumber \\ && \quad + \exp[\beta (J_z/2 - D)] 
[\cosh(\beta R) + (J_z - 2D) \sinh(\beta R)/2R] \Bigr \}. 
\label{eq5d}	   
\end{eqnarray}  
It is worthwhile to mention that the quantities listed above can also be calculated 
from the definition of statistical mean value after performing the transformation
of the relevant quantity into the basis of eigenfunctions, i.e.
by means of: $\langle ... \rangle = \frac{1}{{\mathcal Z}} 
\sum_{\alpha} \langle \Psi_{\alpha}|...| \Psi_{\alpha} \rangle e^{-\beta \lambda_{\alpha}}$ (summation runs over all eigenstates).
This alternative approach is even more general, since it enables to calculate 
the ensemble average for any combination of random spin variables and  
not only for those, which are explicitly involved by the different interaction parameters in the total spin Hamiltonian. As an example, we have evaluated within 
this approach the transverse component of quadrupolar moment, the quantity, 
which cannot be calculated by differentiating the Gibbs free energy: 
\begin{eqnarray}
C_{11}^{xx} &=& \langle (\hat S_1^x)^2 \rangle = \langle (\hat S_1^y)^2 \rangle =
\frac{1}{\mathcal Z} \Bigl \{\exp[- \beta (J_z + 2D)] \cosh(2 \beta H) 
\nonumber \\ &+& 3 \exp(- \beta D) \cosh(\beta H) \cosh(\beta J_{xy}) + 
\exp[\beta (J_z - 2D)]/2 \nonumber \\ &+& \exp[\beta (J_z/2 - D)] 
[3 \cosh(\beta R)/2 - (J_z/2 - D) \sinh(\beta R)/2R] \Bigr \}.
\label{eq6}	   
\end{eqnarray} 
It is noticeable that the calculated correlation functions and spatial components of quadrupolar moment provide, in addition to the better understanding 
of the spin order, useful connection to the macroscopically measurable quantities 
such as susceptibility, entropy, specific heat, etc. In fact, by exploiting the fluctuation-dissipation theorem one finds following relations between the 
spatial components of molar susceptibility and respectively, magnetization, correlation functions and quadrupolar moment:
\begin{eqnarray}
\chi_x &=& N_{\mathrm{A}} \frac{g_x^2 \mu_{\mathrm{B}}^2}{k_{\mathrm{B}} T} 
                        (C_{11}^{xx} + C_{12}^{xx}), \qquad
\chi_y = N_{\mathrm{A}} \frac{g_y^2 \mu_{\mathrm{B}}^2}{k_{\mathrm{B}} T} 
                        (C_{11}^{yy} + C_{12}^{yy}), \nonumber \\
\chi_z &=& N_{\mathrm{A}} \frac{g_z^2 \mu_{\mathrm{B}}^2}{k_{\mathrm{B}} T} 
                        (C_{11}^{zz} + C_{12}^{zz} - 2m_z^2).
\label{eq7}	   
\end{eqnarray} 
With regard to earlier results (\ref{eq5a})-(\ref{eq6}), the set of Eqs. (\ref{eq7}) yields the 
closed-form expressions allowing the exact calculation of the longitudinal ($\chi_z$) as well as transverse ($\chi_x, \chi_y$) components of the molar susceptibility ($N_{\mathrm{A}}$ denotes Avogadro's number). 
Of course, similar relations can also be established between the internal energy, entropy or specific heat 
on the one side and the magnetization, correlation functions and quadrupolar moment on the other side. However, these thermodynamic quantities can be readily obtained from the Gibbs free energy with the help
of standard thermodynamic relations and it is therefore meaningless to use such an approach for them.     

\section{Results and discussion}

Before proceeding to the discussion of the most interesting analytical and numerical results, it should be emphasized that the exact results derived in the previous section are rather general, in fact, they hold for any combination of intra-dimer coupling constants $J_{xy}$ and $J_z$ (both of them can be F or AF), arbitrary uniaxial anisotropy $D$ and external magnetic field $B$. For simplicity, we will suppose that both coupling constants are of equal sign, i.e. they are either  ferromagnetic ($J_{xy}, J_z < 0$), or antiferromagnetic ($J_{xy}, J_z > 0$). 

\subsection{Ground-state behavior}

At first, let us focus on the ground-state properties of spin-1 dimer.
Since the ground state is definitely determined by the lowest-energy 
eigenstate, it is just a problem of finding the lowest energy level
among all possible eigenenergies. 

\subsubsection{Zero-field case}

In the zero-field case we deal with a rather simple situation. It can be easily proved that the $|\Psi_{0,0} \rangle$ is always the lowest-energy eigenstate when both coupling constants are AF ($J_{xy}, J_z > 0$). In the consequence of that, the spin arrangement of the spin-1 AF dimer can be characterized with the aid of:
\begin{eqnarray}
C_{12}^{xx} &=& - \frac{2 J_{xy}}{\sqrt{(J_z - 2D)^2 + 8J_{xy}^2}}, \qquad 
C_{11}^{xx} = \frac14 \Bigl(3 - \frac{J_z - 2D}
                                        {\sqrt{(J_z - 2D)^2 + 8 J_{xy}^2}} \Bigr),
\nonumber \\                                        
C_{12}^{zz} &=& - C_{11}^{zz} = - \frac12 \Bigl(1 + \frac{J_z - 2D}
                                        {\sqrt{(J_z - 2D)^2 + 8J_{xy}^2}} \Bigr),
                                        \qquad m_z = 0.
\label{eq8}	   
\end{eqnarray}  
It can be easily understood that the strength of AF spin alignment within different spatial directions depends basically on a competitive influence of $J_z$, $J_{xy}$ and $D$ interaction parameters. As could be expected, the \textit{easy-axis single-ion anisotropy} forces spins to align anti-parallel with respect to each other, 
letting $D \to -\infty$ indeed gives $C_{12}^{zz}=-C_{11}^{zz}=-1$ similarly as it is in the case of spin-1 Ising dimer, i.e. under the assumption of the infinitely strong \textit{easy-axis exchange anisotropy} 
($J_z \to \infty$). On the contrary, there is an obvious distinction between the effect of 
exchange and single-ion anisotropies in the easy-plane regime. The \textit{easy-plane single-ion anisotropy} clearly tries to tend both spins in the $xy$-plane. Apart from $C_{12}^{zz}=C_{11}^{zz}=0$ retrieved in the limit $D \to +\infty$, the $|\Psi_{0,0} \rangle$ eigenvector can serve in evidence of this statement in that it reduces towards a single $| 0,0 \rangle$ state. On the other hand, the population of the 'intrinsic AF'  
$|\pm 1, \mp 1 \rangle$ states does not completely vanish owing to the \textit{easy-plane exchange anisotropy} even if $J_{xy} \to \infty$. Actually, the spin-1 dimers are constrained to remain in the superposition of spin states even under this condition and the $| 0,0 \rangle$ state is occupied with the same probability as both 'intrinsic AF' $|\pm 1, \mp 1 \rangle$ states together. An independent check of this scenario also provides $C_{12}^{zz}=-C_{11}^{zz}=-0.5$ retrieved in the $J_{xy} \to \infty$ limit.

Surprisingly, the situation becomes a bit more involved when assuming a ferromagnetic version of the spin-1 dimer ($J_{xy}, J_z < 0$). The doubly degenerate energy level that corresponds to both eigenvectors $|\Psi_{2, \pm 2} \rangle$ with a parallel spin alignment (spins are oriented either upwards or downwards) 
is, namely, the lowest-lying only in a certain subspace of interaction parameters. Really, the \textit{easy-plane anisotropy} may induce a level crossing between the doublet of F eigenstates $|\Psi_{2,\pm 2} \rangle$ and respectively, the  eigenstate $|\Psi_{2, 0} \rangle$ which can also be characterized by set of Eqs. (\ref{eq8}). It can be realized that the considered level crossing occurs on behalf of the competition between $J_z$, $D$ and $J_{xy}$ when: 
\begin{eqnarray}
-J_z^{lc} = |J_z^{lc}| = D + \sqrt{D^2 + J_{xy}^2}, \qquad 
(\mbox{for} \: \: J_z < 0 \: \: \mbox{only}).
\label{eq9}	   
\end{eqnarray}
The lowest-energy level is the doublet of $|\Psi_{2,\pm 2} \rangle$ states if $|J_z|>|J_z^{lc}|$ is 
satisfied, otherwise the non-degenerate $|\Psi_{2, 0} \rangle$ state possesses the lowest energy. 
As far as the single-ion (exchange) anisotropy is concerned by itself only, the value matching 
the level crossing being $-J_z^{lc}=2D$ ($J_{z}^{lc}=J_{xy}$). Accordingly, it can be easily 
understood that an appearance of the 'symmetric' AF eigenstate $|\Psi_{2, 0} \rangle$ in the 
region of F coupling constants ($J_{xy}, J_z < 0$) is strongly reinforced when the easy-plane 
single-ion anisotropy acts simultaneously with the easy-plane exchange anisotropy. However, 
it should be also mentionedthat the $|\Psi_{2, 0} \rangle$ state is energetically favored before 
the doublet of F states $|\Psi_{2,\pm 2} \rangle$ provided that the easy-axis exchange anisotropy 
is accompanied with much stronger easy-plane single-ion anisotropy or naturally, in the reverse case.

\subsubsection{Effect of external field - the magnetization process}

In this part, we will explore the effect of external magnetic field on the character of ground-state spin arrangement. To simplify further manipulations, we will rescale all interaction parameters with respect to the exchange coupling $|J_{xy}|$ 
by introducing the following set of normalized parameters:
$\lambda = J_z/|J_{xy}|; \: \: d = D/|J_{xy}|; \: \: h = H/|J_{xy}| 
= g_z \mu_{\mathrm{B}} B/|J_{xy}|; \: \:
t = k_{\mathrm{B}}T/|J_{xy}|$.

For illustration, Fig. 1 displays the regions corresponding to the lowest-energy eigenstates in the $d-h$ plane for several values of the exchange anisotropy $\lambda$. Upper (lower) panel fits the situation by considering the spin-1 AF (F) dimer; the left, central and right panels serve for a comparison of the easy-plane, isotropic and easy-axis interaction regimes. Apart from the  $|\Psi_{0,0} \rangle$ and $|\Psi_{2,0} \rangle$ eigenstates, which have already been described by means of Eqs. (\ref{eq8}), 
the other allowable zero-point eigenstates can be characterized as:
\begin{eqnarray}
|\Psi_{1,1} \rangle \! : \: 
C_{12}^{xx} &=& -0.5, \: \:  C_{11}^{xx} = 0.75, \: \:  C_{12}^{zz} = 0.0, \: \:  C_{11}^{zz} = 0.5, \: \: m_z = 0.5;  \label{eq10a}	 \\
|\Psi_{2,1} \rangle \! : \: 
C_{12}^{xx} &=& 0.5, \: \: \: \: \: C_{11}^{xx} = 0.75, \: \:  C_{12}^{zz} = 0.0, 
\: \: C_{11}^{zz} = 0.5, \: \: m_z = 0.5; \label{eq10b}	  \\
|\Psi_{2,2} \rangle \! : \:  
C_{12}^{xx} &=& 0.0, \: \: \: \: \: C_{11}^{xx} = 0.5, \: \: \: \: C_{12}^{zz} = 1.0, \: \: C_{11}^{zz} = 1.0, \: \: m_z = 1.0.  
\label{eq10c}	   
\end{eqnarray}  
Further, the boundaries separating the lowest-energy eigenstates are given by:
\begin{eqnarray}
&& |\Psi_{0,0} \rangle (|\Psi_{2,0} \rangle) - |\Psi_{1,1} \rangle (|\Psi_{2,1} \rangle) \! : \: \quad
h = \lambda/2 - 1 + \sqrt{(d - \lambda/2)^2 + 2}; \label{eq11a} \\
&& |\Psi_{1,1} \rangle (|\Psi_{2,1} \rangle) - |\Psi_{2,2} \rangle \! : \: \quad
h = \lambda + d + 1; \label{eq11b} \\
&& |\Psi_{0,0} \rangle (|\Psi_{2,0} \rangle) - |\Psi_{2,2} \rangle \! : \: \quad 
h = 3\lambda/4 - d/2 + \sqrt{(d - \lambda/2)^2 + 2}/2; 
\label{eq11c}	   
\end{eqnarray}   
and for completeness, we also write in a parametric form the coordinates of special 
points at which all three boundary lines cross each other:
\begin{eqnarray}
\lambda > -2; \qquad d_t = -(\lambda + 1)/(\lambda + 2); \qquad
                     h_t = (\lambda + 1)^2/(\lambda + 2). 
\label{eq12}	   
\end{eqnarray}   
Apparently, the zero-temperature magnetization curves are obtained from Fig. 1 when
moving along the vertical $h$-axis. If doing so, one gains a stepwise magnetization curve with the abrupt change(s) of magnetization at crossing field(s) given by Eqs. (\ref{eq11a})-(\ref{eq11c}). It is quite obvious that there are at best two distinct types of magnetization curves for the spin-1 AF dimer. Firstly, one encounters a single-step magnetization curve with a direct field-induced transition from the $|\Psi_{0,0} \rangle$ state to the $|\Psi_{2,2} \rangle$ state with saturated magnetization. Secondly, there also appears a two-step magnetization curve with an intermediate plateau at half of the saturation magnetization in between the above mentioned $|\Psi_{0,0} \rangle$ and $|\Psi_{2,2} \rangle$ states. Evidently, 
a presence of the intermediate plateau reflects the energetical stability of the $|\Psi_{1,1} \rangle$ state in the range of moderate fields. As could be expected, the easy-axis (easy-plane) anisotropy prefers the single-step (two-step) magnetization curves no matter whether it is caused by the single-ion anisotropy or exchange anisotropy term. Finally, it is noteworthy that the purely isotropic spin-1 AF dimer 
($\lambda= 1, d = 0$) shows the double-plateau magnetization curve.

As far as the spin-1 F dimer is concerned, the stepwise magnetization curves appear just as the claim $|J_z|<|J_z^{lc}|$ is fulfilled because of the strong enough easy-plane anisotropy. Despite the magnetization curves seem in such a case nearly the same as those of the spin-1 AF dimer, there are nevertheless some principal differences. At first, the zero-field as well as intermediate-plateau states do not occur owing to a presence of the $|\Psi_{0,0} \rangle$ and $|\Psi_{1,1} \rangle$ states, but on behalf of the $|\Psi_{2,0} \rangle$ and $|\Psi_{2,1} \rangle$ states. Though these eigenfunctions differ from each other just in the symmetry, this change gives rise to a change of the character of spin correlations in the $xy$-plane (see Eqs. (\ref{eq8}), (\ref{eq10a}) and (\ref{eq10b})). Another important difference emerges when considering the combination of the easy-axis exchange anisotropy and easy-plane single-ion anisotropy. As the easy-axis exchange anisotropy strengthens, the boundary lines separating the intermediate-plateau state $|\Psi_{2,1} \rangle$ merge together until the plateau state wholly disappears from the magnetization curve below $\lambda \leq -2$. This behavior is in a sharp contrast with what is observed in the spin-1 AF dimer, where the plateau state $|\Psi_{1,1} \rangle$ does not vanish due to the easy-plane single-ion anisotropy nothwithstanding of the effect of easy-axis exchange anisotropy. 

To conclude our analysis of the ground state, we depict in Fig. 2 the Zeeman splitting of energy levels in the spin-1 AF dimer by selecting several combinations of the anisotropy parameters $d$ and $\lambda$. Fig. 2a shows a well-known picture matching the situation in the isotropic spin-1 dimer; the energy spectrum consists 
of the singlet ($|\Psi_{0,0} \rangle$), triplet ($|\Psi_{1,0} \rangle$, $|\Psi_{1,\pm 1} \rangle$) and quintet ($|\Psi_{2,0} \rangle$, $|\Psi_{2,\pm 1} \rangle$, $|\Psi_{2,\pm 2} \rangle$) energy levels and the level degeneracy is entirely 
lifted by any non-zero external field. To provide a deeper insight, Fig. 2b (Fig. 2c) illustrates the effect of easy-plane (easy-axis) single-ion anisotropy and Fig. 2d (Fig. 2e) the effect of easy-plane (easy-axis) exchange anisotropy on the overall trends in the energy spectrum. As one can see, both anisotropy parameters are responsible for the zero-field splitting: the triplet level is consequently splitted into the doublet $|\Psi_{1,\pm 1} \rangle$ and the singlet $|\Psi_{1,0} \rangle$ level, whereas the quintet level splits into two doublets $|\Psi_{2,\pm 1} \rangle$, $|\Psi_{2,\pm 2} \rangle$ and one singlet $|\Psi_{2,0} \rangle$ level. Notice that the zero-field splitting induced by the exchange anisotropy and the one originating from the single-ion anisotropy differ from each other even if both anisotropies are assumed to be of the same 
easy-axis (easy-plane) type. Apparently, the aforementioned difference rests 
in the splitting of the quintet level, while the triplet level is, on the contrary, splitted in the qualitatively same manner. Finally, it should be pointed out that by reverting the sign of vertical 
(energy) axis, one obtains the energy spectrum for the spin-1 F dimer with the interaction parameters 
given by the transformation $(\lambda, d) \to (-\lambda, -d)$.

\subsection{Finite-temperature behavior}

Now, we shall pay our attention to the finite-temperature behavior of the magnetization and susceptibility, the quantities, which characterize a response of the magnetic system with respect to the external field. 
Both possible kinds of magnetization curves (single-step as well as two-step) 
are illustrated in Fig. 3. It is necessary to emphasize that there are no real jumps in the magnetization 
vs. field dependence at any finite temperature and that the true discontinuities occur merely at zero temperature as the external field reaches some of its level-crossing values (\ref{eq11a})-(\ref{eq11c}). 
At finite but sufficiently low temperatures, however, an extremely steep but continuous increase of the magnetization observable in the vicinity of the level-crossing fields strongly resembles the ground-state magnetization jumps. It is easy to understand from Fig. 3 that the abrupt change of magnetization is gradually 
smeared out by the thermal fluctuations as the temperature raises and in the consequence of that, 
the magnetization curve acquires at sufficiently high temperatures a typical magnetization profile of the Langevin-Brillouin type. It is worthy to mention that the depicted magnetization curves render the temperature effect on the overall trends in the magnetization profile even if $\lambda \neq 1$. 

Furthermore, the temperature variations of the longitudinal ($\chi_z$) and transverse ($\chi_{x}$) susceptibilities are displayed in Fig. 4 for several combinations of the anisotropy parameters 
$\lambda$ and $d$. As one can see from Fig. 4a, a round maximum in the $t - \chi_z$ dependence 
flattens and shifts towards higher temperatures when the easy-plane single-ion anisotropy strengthens. Conversely, the height of maximum is almost independent of the easy-axis single-ion anisotropy strength
and there is only a slight shift of the maximum position by changing this type of magnetic anisotropy 
(Fig. 4b). As far as the thermal variation of $\chi_{x}$ is concerned, the transverse susceptibility 
diverges by approaching the zero temperature whenever $d \neq 0$ and/or $\lambda \neq 1$, i.e. whenever 
there is at least small magnetic anisotropy. It is quite apparent from Figs. 4c-d that there is 
a quantitative rather than qualitative difference between the effect of easy-plane and 
easy-axis single-ion anisotropies on the thermal dependence of transverse susceptibility. At relatively weak single-ion anisotropies ($|d| < 0.75$), the $t-\chi_{x}$ dependence exhibits, in addition to the familiar round maximum, the successive minimum before showing the final low-temperature divergence under the temperature suppression. The round minimum in the $t - \chi_{x}$ dependence gradually flattens as the single-ion anisotropy strengthens and the transverse susceptibility becomes a monotonical decreasing function of temperature above $|d| > 0.75$. Finally, we just quote that the addition of the exchange anisotropy has
a very similar effect on the thermal dependences of longitudinal and transverse susceptibilities as the single-ion anisotropy.

\section{Magnetic behavior of homodinuclear nickel(II) complex \newline           
[Ni$_2$(Medpt)$_2$($\mu$-ox)(H$_2$O)$_2$](ClO$_4$)$_2$.2H$_2$O (NAOC)}

In this part, the obtained theoretical results will be confronted with the relevant experimental measurements 
performed on the single-crystal sample of [Ni$_2$(Medpt)$_2$($\mu$-ox)(H$_2$O)$_2$](ClO$_4$)$_2$.2H$_2$O (hereafter abbreviated as NAOC), which is regarded as a typical representative of the spin-1 dimeric 
compound \cite{EVR}. A schematic view on the dinuclear unit, relative positions of non-ligating 
perchlorate counter-anion and water molecules are presented in Fig. 5. Both octahedrally 
coordinated nickel(II) atoms are linked via bis-chelating oxalato group and the rest 
of their coordination sphere is occupied by the blocking tridentate amine (Medpt) and ligating 
water molecule. Accordingly, the single-crystal sample of NAOC consists of an assembly of discrete 
dinuclear entities, which are connected through the hydrogen bonds and/or van der Waals forces
to the molecular crystal. It can be viewed from Fig. 5 that the NAOC molecule is centrosymmetric with 
an inversion center at the midpoint of the bridging oxalato group. Thereupon, the antisymmetric Dzyaloshinski-Moriya interaction \cite{DM} (a first-order effect of spin-orbit coupling and exchange interaction) should be entirely neglected as a possible source of the magnetic anisotropy in the NAOC. The exchange anisotropy, i.e. the anisotropy in a symmetric (pseudodipolar) exchange interaction, could be then anticipated to be the third strongest source of magnetic anisotropy beyond the single-ion anisotropy and the $g$-factor anisotropy.

Before proceeding to a comparison of the theoretical and experimental results it should be noted here 
that the magnetic measurements performed on the NAOC sample have been primarily reported by some of the present authors in the earlier publications \cite{Narumi}-\cite{ESR} to which the interested reader is referred for a closer experimental details. Notice that both magnetization as well as susceptibility data reported on previously imply strongly anisotropic magnetic behavior \cite{Narumi} and the marked anisotropy manifests itself also in the peculiar angular dependence of level-crossing fields (see Fig. 1 of Ref. \cite{ESR}). By taking into account the results of Ref. \cite{ESR}, we choose the crystallographic $c^{\star}$-axis as the principal axis because $c^{\star}$ being the easy magnetization axis. 
The crystallographic $a$-, $b$- and $c^{\star}$-axis are consequently defined to be the $x$-, $y$- 
and $z$-axis of the effective spin Hamiltonian. Finally, we should also remark that the high-field magnetization data used in our further analysis are identical with those reported on earlier \cite{Narumi}, 
while the original susceptibility data were found erroneous and we report here a revised set of data for them.
 
Magnetization normalized with respect to its saturation value are plotted in Fig. 6 against 
the external magnetic field applied along the $c^{\star}$- and $a$-axis. Unfortunately, there 
does not exist an exact analytical expression for the magnetization of spin-1 dimer when the external 
field is applied perpendicular to the principal axis. Adopting the least-square fitting based 
upon a minimalisation of the function $R = \sum (A_i^{cal} - A_i^{obs})^2 / \sum (A_i^{obs})^2$, 
both displayed magnetization curves were therefore fitted through the relation (\ref{eq5}) derived 
for the magnetization along the principal direction. At first, we decided to neglect the exchange 
anisotropy by keeping both exchange parameters equal to each other ($J_{xy}=J_z=J$) and only the gyromagnetic ratio $g$ together with the single-ion anisotropy parameter $D$ were allowed to vary within different spatial directions. Under these conditions, the best simultaneous fit of all magnetization data is obtained for the following set of parameters (Fig. 6): $J/k_{\rm B} = 30.66$K and $D_{c^{\star}}/k_{\rm B} = -12.48$K, $g_{c^{\star}} = 2.28$ ($R_{c^{\star}} = 0.0019$) for $c^{\star}$-axis and respectively, 
$D_a/k_{\rm B} = 4.91$K, $g_a = 2.24$ ($R_a = 0.0067$) for $a$-axis. In agreement with 
our expectations, the magnetization fit is substantially better for the $c^{\star}$-axis than for 
the $a$-axis ($R_{c^{\star}}<R_a$) and the anisotropy parameter $D$ has an opposite sign for the 
easy ($c^{\star}$) and hard ($a$) magnetization axis.

On account of a relatively strong easy-axis single-ion anisotropy, the intermediate plateau shrinks 
to a very narrow field range when the external field is applied along the easy magnetization axis (Fig. 6a). 
Moreover, it is interesting to note that the obtained single-ion anisotropy strength 
$D_{c^{\star}}/J \approx  -0.41$ is rather close to the boundary value $D_t/J = - \frac23$ below which the intermediate plateau should completely disappear from the magnetization curve according to the theoretical prediction based on Eq. (\ref{eq12}). The inverse effect is observed, however, when applying the external field normal to the easy magnetization axis (Fig. 6b). In this case, the field range corresponding to the intermediate plateau largely extends not only owing to a much higher saturation field, but also due to a slightly lower field required for attaining the plateau state (see the insert of Fig. 6b). 
Finally, it is worthy to note that we found a strong evidence that the magnetic behavior observed in the NAOC can be undoubtedly attributed purely to the single-ion anisotropy effect. The exchange anisotropy is unlikely to significantly contamine the magnetochemistry of the NAOC, actually, the improvement of $R_a$ and $R_c^{\star}$ quality factors accomplished by the inclusion of this anisotropic term does not exceed 0.5\%.  

The temperature dependences of susceptibility measured along the $a$-, $b$- and $c^{\star}$-axis 
are compared in Fig. 7 with the relevant theoretical prediction for the longitudinal and transverse susceptibility. Notice that the experimental susceptibilities were corrected for the diamagnetic contribution according to Pascal tables. Under the restriction to the isotropic intra-dimer interaction 
$J$, the best simultaneous fit of all susceptibility data was obtained for this unique set of interaction parameters: $J/k_{\rm B} = 30.89$K, $D/k_{\rm B} = -5.61$K, $g_a = 2.34$ ($R_a = 0.0028$), $g_b = 2.30$ 
($R_b = 0.0019$) and $g_{c^{\star}} = 2.37$ ($R_{c^{\star}} = 0.0050$). Evidently, there is a rather reasonable accord between the values of $J$, $g_a$, $g_b$ and $g_{c^{\star}}$ parameters listed in the 
fitting set of magnetization and susceptibility, as a matter of fact, the latter values are only slightly enhanced with respect to the former ones. The most significant discrepancy thus rest in the strength of uniaxial single-ion anisotropy $D$, which is more than twice weaker for the susceptibility fit than for the 
magnetization fit.

There are strong structural indications that the inconsistency in determining the uniaxial zero-field 
splitting parameter $D$ can be related to higher-order anisotropy terms, which were entirely neglected 
in our study. The ligand asymmetry in the equatorial plane of Ni$^{2+}$ ions (see Fig. 5) indeed implies 
a possibly non-negligible biaxial single-ion anisotropy and the numerical calculations involving this anisotropic term strongly support this concept \cite{ESR}. Unfortunately, the influence of biaxial 
anisotropy was out of scope of the present study, since this term prevents an exact analytical treatment 
used here. Further, it turns out that the monotonical decrease of susceptibility observed upon cooling 
ceases below $5K$ independently of the spatial direction, what indicates a small fraction of non-coupled 
nickel(II) impurities to be present in the sample. However, the best simultaneous fit of all susceptibility data which accounts also for the paramagnetic impurities considerably reduces just an error in 
the quality factor $R_{c^{\star}}$ (from $0.0050$ down to $0.0019$), while the other quality factors 
and fitting parameters remain by the estimated concentration of impurities ($p \approx 0.34 \%$) 
almost unaltered. The most substantial deviations between the theoretical and experimental curves 
thus appear in the high-temperature region, where the theoretical curves lie slightly below the 
experimental ones.

\section{Conclusion}

In the present article, the particular attention has been devoted to a comparative study of the uniaxial single-ion anisotropy and exchange anisotropy as possible sources of the magnetic anisotropy in the 
spin-1 dimer. Exact analytical results for the Gibbs free energy, longitudinal magnetization, correlation functions, quadrupolar moment, longitudinal and transverse susceptibility were derived and on the basis of these results, a simple derivation of other quantities such as internal energy, entropy, or specific heat, 
can easily be accomplished. It is worthy to notice that all aforementioned results can be rather straightforwardly extended to account for an effect of the inter-molecular (inter-dimer) interaction 
treated within the mean-field approximation. 

The most important result stemming from our study consists in the derivation of exact closed-form 
expressions for the longitudinal and transverse susceptibilities by means of the fluctuation-dissipation theorem, which relates these quantities to the magnetization, correlation function and quadrupolar moment. 
To the best of our knowledge, we are not aware of any work providing an exact closed-form formula for the transverse susceptibility of the anisotropic spin-1 dimer. In addition, we found a convincing evidence that the transverse susceptibility of spin-1 dimer diverges as $T \to 0$ whenever there is an arbitrary but non-zero magnetic anisotropy. Finally, the accurate condition allocating the 'symmetric' AF ground state of spin-1 F dimer was determined (\ref{eq9}) and its physical background was discussed in detail.

As far as the single-ion anisotropy and exchange anisotropy effects are compared, the field dependence of magnetization as well as the temperature dependence of susceptibility are influenced by these anisotropic terms in a qualitatively same manner. It is therefore rather intricate to find out, sometimes, whether the overall magnetic anisotropy is attributable purely to one source, or arises as a sum of more contributions. 
It should be mentioned here, however, the high-field ESR measurement provides a very efficient alternative tool that enables to distinguish between the different magnetic anisotropies, since it reflects the overall energy spectrum depending qualitatively on the type of magnetic anisotropy as one can understand from the 
set of eigenenergies (\ref{eqe}).

The obtained theoretical results were also compared with the available experimental data of the NAOC, which is regarded to be a typical representative of the spin-1 AF dimer compound. There is a strong evidence that the zero-field splitting induced by the uniaxial single-ion anisotropy is by far the most dominating phenomenon which determines the overall magnetic anisotropy in the NAOC compound. Nevertheless, it is interesting to mention that the strength of uniaxial single-ion anisotropy evaluated from the magnetization fit is quite close to the boundary value below which the system should exhibit a striking spatially-dependent magnetization process. Subsequently, one should encounter the single-step magnetization curve when applying the external field along the easy magnetization axis, while the two-step magnetization curve should be expected to occur when the field is applied perpendiculary to it. Although the design of molecule-based magnetic materials 
with a tunable strength of the single-ion anisotropy is far from a routine target at present, a large amount 
of oxalato-bridged dinuclear nickel complexes \cite{ox} supports our hope that this family of compounds will provide a suitable candidate for displaying such an interesting phenomenon. 

\ack{This work was financially supported under the grants VEGA 1/2009/05 and APVT 20-005204.}

\newpage

\begin{large}
\textbf{Figure captions}
\end{large}

\begin{itemize}

\item [Fig. 1]
The lowest-energy eigenstates in the $d-h$ plane for several values of exchange anisotropy $\lambda$. 
Upper (lower) panels fits the situation in the spin-1 AF (F) dimer, the left, central and right panels 
serve for a comparison of the easy-plane, isotropic and easy-axis interaction regimes, respectively.  

\item [Fig. 2]
Zeeman's splitting of the complete energy spectrum shown for several combinations of the anisotropy parameters 
$\lambda$ and $d$: a) $\lambda = 1.0, d = 0.0$; b) $\lambda = 1.0, d = 0.5$; c) $\lambda = 1.0, d = -0.5$;
d) $\lambda = 0.5, d = 0.0$; and e) $\lambda = 2.0, d = 0.0$. 
For clarity, the indices of eigenvectors are shown only. 

\item [Fig. 3]
The typical examples of two-step and single-step magnetization curves displayed at various dimensionless temperatures $t = 0.00, 0.05, 0.10, 0.20, 0.50$ in ascending order along the direction of arrows. 
The parameters used are $\lambda = 1.0, d = 0.50$ (Fig. 3a) and $\lambda = 1.0, d = - 0.75$ (Fig. 3b), 
the magnetization is scaled in the $g_z \mu_B$ units.
      
\item [Fig. 4]
The thermal dependence of the zero-field susceptibility scaled in the 
$\mbox{N}_{{\rm A}} (g \mu_{{\rm B}})^2 / k_{\rm B}$ 
units when the parameter $\lambda$ is fixed and the parameter $d$ varies. Upper (lower) panel 
shows the longitudinal (transverse) susceptibility, the left (right) panels illustrates the situation 
under the easy-plane (easy-axis) single-ion anisotropy.

\item [Fig. 5]
Schematic representation of the NAOC dinuclear unit, non-ligating perchlorate counter anion and 
water molecules. The NAOC molecule has an inversion center at the midpoint of the oxalato bridging group. 

\item [Fig. 6]
The high-field magnetization curves measured in the pulsed magnetic field along the $a$- and $c^{\star}$-axis 
at very low temperature $T = 1.3K$. The best simultaneous fit, determined as described in the text, was obtained for the following set of fitting parameters: $J/k_{\rm B} = 30.66K$ and $D_{c^{\star}}/k_{\rm B} = -12.48K$, $g_{c^{\star}} = 2.28$ ($R_{c^{\star}} = 0.0019$) for $c^{\star}$-axis and respectively, $D_a/k_{\rm B} = 4.91K$, $g_a = 2.24$ ($R_a = 0.0067$) for $a$-axis. The solid (dashed) lines show the experimental (theoretical) curves, the insert depicts both experimental magnetization curves in the same scale. 
 
\item [Fig. 7]
The thermal variations of magnetic susceptibility measured along the $a$-, $b$- and $c^{\star}$-axis 
plotted together with the best simultaneous fit obtained for: $J/k_{\rm B} = 30.89K$, $D/k_{\rm B} = -5.61K$, $g_a = 2.34$ ($R_a = 0.0019$), $g_b = 2.30$ ($R_b = 0.0021$) and $g_{c^{\star}} = 2.37$ ($R_{c^{\star}} = 0.0050$). For clarity, the theoretical predictions for the longitudinal and transverse susceptibility representing, respectively, the susceptibility measured along the $c^{\star}$- and $b$-axis 
are shown only. 

\end{itemize}
\end{document}